# Adapting Software Quality Models: Practical Challenges, Approach, and First Empirical Results

Michael Kläs, Constanza Lampasona
Fraunhofer Institute for Experimental Software Engineering
Fraunhofer-Platz 1, 67663 Kaiserslautern, Germany
{michael.klaes, constanza.lampasona}@iese.fraunhofer.de

Jürgen Münch
University of Helsinki, Department of Computer Science
Gustaf Hällströmin katu 2b, 00014 Helsinki, Finnland
juergen.muench@cs.helsinki.fi

*Abstract*—Measuring and evaluating software quality has become a fundamental task. Many models have been proposed to support stakeholders in dealing with software quality. However, in most cases, quality models do not fit perfectly for the target application context. Since approaches for efficiently adapting quality models are largely missing, many quality models in practice are built from scratch or reuse only high-level concepts of existing models. We present a tool-supported approach for the efficient adaptation of quality models. An initial empirical investigation indicates that the quality models obtained applying the proposed approach are considerably more consistently and appropriately adapted than those obtained following an ad-hoc approach. Further, we could observe that model adaptation is significantly more efficient (~factor 8) when using this approach.

*Keywords-software quality; customizing quality models; meta-model; tailoring process; empirical study*

## I. INTRODUCTION

Nowadays, the definition and evaluation of software quality is a fundamental task for many organizations: Objective statements about software quality are needed, i.e., for defining and fulfilling software-related contracts, for controlling and adjusting development and quality assurance processes, or for managing quality-related risks. Many organizations refer to so-called quality models (QMs) when addressing these issues.

A plethora of software QMs and quality modeling approaches intended to support product quality stakeholders in dealing with software quality have been developed over the past decades. Most of them can be assigned to one of two strategies for modeling software quality [17], namely fixed-model approaches, e.g., ISO9126 [14], and define-your-own-model approaches, e.g., GQM [3]. The former usually specify a prescriptive set of quality characteristics or metrics, whereas the latter use methods to guide the experts in the derivation of customized QMs. The fixed models are often very abstract and therefore not directly applicable [28][29] or their applicability is limited to contexts that are very similar to the one in which the model was developed. In contrast, define-your-own-model approaches can be applied to obtain QMs fitting one's own needs but require the labor-intensive involvement of experienced experts, who are typically one of the most limited resources in a company.

An initial step towards overcoming the gap between fixed-model and define-your-own-model approaches and making the modeling of customized QMs more efficient can be seen in choosing an existing model that is most appropriate for one's own needs and reusing parts of the model during the application of a define-your-own-model approach as suggested, e.g., by [16]. However, this implies that an appropriate model must be identified, which, considering the high number of existing QMs, is a non-trivial task for practitioners. In previous work, we have addressed this challenge by developing a classification schema for QMs that can be used to identify appropriate QMs in a goal-oriented way [18] and by providing classifications for about 80 QMs.

Still, identifying a (partially) reusable QM is only the first step. In most cases, the identified QM does not fit perfectly to the target application context and needs to be adapted in a next step. For instance, irrelevant parts have to be removed, other parts require some modification, and missing parts have to be created. Although such customization is a complex, fault-prone, and effort-intensive task for real-world QMs, work dealing with the efficient adaptation of software QMs is largely missing. This is especially remarkable since the adaptation of a QM is not a one-time task. It is performed when QMs are defined and introduced, but also as part of ongoing maintenance to keep the applied QMs consistent with the needs of the organization and thus have a sustainable instrument for quality management.

Based on the requirements stated by practitioners and scientists in [27], we condensed three major requirements with respect to a QM adaptation approach:

(R1) *Correctness* – An adapted QM must be syntactically correct in that it remains conformant to its underlying structure and a set of defined consistency rules.

(R2) *Appropriateness* – The adaption of a QM should be driven by organizational needs and capabilities. In particular, organization-specific and project-specific software quality objectives should be considered.

(R3) *Efficiency* – This requirement is concerned with the overhead (e.g., personnel, time, and budget) needed for adapting a QM. Acceptable overhead would differ depending on the organizational level (e.g., more overhead will be allowed for adapting a QM at the level of the whole organization, where such adaptation has a larger scope and is performed relatively rarely).

In the following, we present an adaptation method addressing the listed requirements. One major challenge regarding the definition of such a QM adaptation method is to make it as independent as possible of a particular model and type of adaptation, i.e., to define a set of adaptation rules that will be universally applicable to any model and adaptation scenario. However, making no assumptions about the underlying structure of the QMs that should be adapted would avoid the operationalization of the adaptation method; therefore, we assume that the adapted QM should conform to the structure and consistency rules defined by the Quamoco quality meta-model [19]. This meta-model was developed in a joint effort by academic and industrial partners in a publicly funded project and addresses all conceptual elements recommended in [18] for specifying and assessing software quality. Recent empirical evaluations have shown that the meta-model is general enough to describe many different QMs applied in practice [19] and specific enough to define QMs that can be used for valid product quality evaluations [21].

This paper consolidates and extends the work presented by the authors at the SQMB workshops in 2010 [20] and 2011 [23]. In the following, we provide an overview of related work in QM adaptation and take a brief look at adaptation approaches in related areas. Then, we give an introduction to the adaptation method and the underlying meta-model. Next, we present an initial study we performed to evaluate the approach and discuss the study's findings. Finally, we summarize our current work and sketch planned research directions.

## II. RELATED WORK

Most of the literature on adapting QMs deals with adaptations of the QM proposed by the ISO9126 standard [14]. Many authors concentrate their work on extending the ISO9126 QM with quality attributes, such as in the adaptations in [24] and [8]. Behkamal et al. [5] add domain-specific quality characteristic to a model for B2B applications; Andreou and Tziakouris [2] do so for component-based software development, and Calero et al. [8] for eBanking applications. Unfortunately, these specific adaptations focus on the resulting adapted QMs and not on a reproducible customization process.

Another practice for adapting QMs consists of using define-your-own-model tools to refine specific models. Andersson and Eriksson [1], e.g., present a process for the construction of a QM founded on a basic QM with existing metrics (SOLE QM). They illustrate how to customize the model to the specific needs of an organization, including how to identify quality factors and mapping them down to metrics. Their model [9] has the factor-criteria-metric [24] structure. Bianchi et al. [6] used GQM to refine a specific model. They focus on QM reuse, namely, which changes can be requested when a QM is reused, how to verify that the changes made in the reused QM keep it suitable for its goals, and which are the side effects on the QM caused by changing the metrics. Khaddaj and Horgan [15] use as input the adaptable QM (ADEQUATE), which provides a set of standard quality factors. Any decisions are made by experts. Franch and Carvallo [12] present a general process for building an ISO9126-based QM. These customizations focus on a specific QM or do not assume a common underlying structure and therefore their adaptation guidelines are only rough and difficult to operationalize for adaptation in practices. Plösch et al. [26] present a tool-supported approach to adapting QMs focused on code evaluation. They tailor a set of rules provided by static code analysis tools based on a set of criteria. Although the scope and structural complexity of their model is limited compared to more universal QMs, the general idea of providing a well detailed and comprehensive model that is primarily reduced during adaptation following certain criteria seems promising in terms of ensuring efficient adaptation.

For the refinement of our solution, we also considered concepts related to software process adaptation and studied their transferability to software product QMs. Software process tailoring emerged from the need to reuse process definitions, a motivation analogical to that of software QM adaptation. Software processes had to be developed from scratch or projects had to be forced to fit prescriptive processes. This problem has been managed by defining standard software processes and tailoring them to obtain project-specific processes in accordance with project needs [13], project goals, and context characteristics [4].

Budlong et al. [7] propose steps for adapting a standard software process for use in a specific project. The first step involves identifying project characteristics (size, complexity, formality, and control). Afterwards, relevant building blocks are selected from an inventory and subsequently tailored to the project characteristics. Fitzgerald et al. [11][10] describe development process components across three levels: industrial level, organizational level, and project level. The tailoring process consists of refining these components first from the industrial to the organizational level. Then, the organizational process can be customized for individual projects. The German *V-Modell* is a tailorable process model [22] based on a meta-model that defines a language for the *V-Modell* and supports an adaptation on an organizational and project-specific level.

Münch [25] proposes context-oriented alignment of process patterns, an abstract description of one or more software development processes, to project goals and project environment characteristics. Fundamental challenges for successful process pattern adaptation are the identification of the necessary initial changes and the consistent performance of concrete consequential changes.

## III. QUALITY MODEL ADAPTATION METHOD

Our approach makes use of many adaptation concepts, which are scattered across different application domains and not necessarily focused on tailoring QMs. The main concepts in our approach are that (1) all the QMs used and produced respect the same general structure, given by a *quality meta-model*, which is required to provide specific rules and automation; (2) adaptations can be performed to refine models on different levels, e.g., for the *organization* or for *individual projects*, reducing the adaptation effort by increasing the reuse potential; (3) the *goal and the application context* of the adapted model are explicitly considered to obtain an appropriate model; (4) a *preliminary adaptation* (tailoring) is performed based on the goal and context to simplify the remaining adaptation work; and (5) *task-specific guidelines* are provided to improve the consistency and completeness of the performed adaptation.

## A. The Quamoco Quality Meta-Model

In order to comprehend the adaptation method, a rough understanding of the structuring principals assumed for the adapted QMs is required. Based on the Quamoco meta-model, a QM can be logically separated into two parts: a mandatory *specification* part, where quality is described qualitatively, and an optional *evaluation* part, which is needed if quality assessments are to be performed (Fig 1).

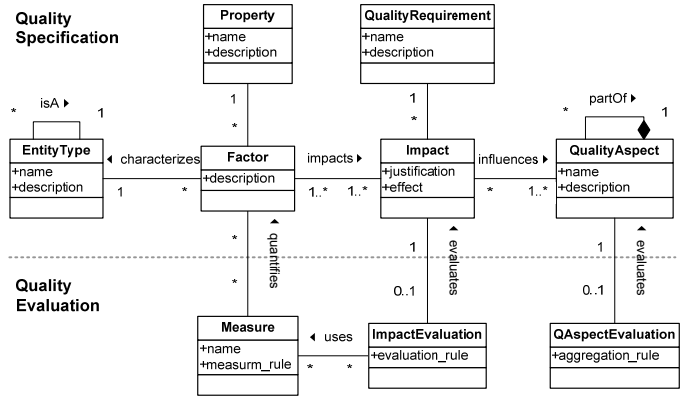

Figure 1. Quamoco meta-model for software QMs

Six types of elements are used to specify quality: A tree of *Quality Aspects* provides information on the quality focused on in the model, e.g., the "maintainability" of a product, and can be used to decompose it into sub-aspects (e.g., analyzability, testability, etc.). *Entity Types* represent the different classes of elements that are part of the software product, e.g., specification, source code, functions, identifiers. Elements of a specific entity type typically exhibit certain *Properties* (such as consistency, conciseness, or redundancy) that influence one or more quality aspects. Hence, *Factors* combine an entity type and a measurable property, e.g., "consistency of identifiers" or "understandability of source code comments". *Impacts* specify which factors influence which quality aspects, their assumed effect (positive or negative), and provide a justification for the relationship. For example, a good "consistency of identifiers" has a "positive" impact on the "maintainability" of the product. Finally, *Quality Requirements* can be used to group a set of related impacts.

For the evaluation of quality, three additional constructs are provided: *Measures* define methods for quantifying factors using a certain measurement rule and scale. Examples of measures are "number of incompletely documented use cases" or "number of architecture violations". *Impact Evaluations* provide rules for determining the impact of a factor on a quality aspect using the data collected for associated measures. This means that the impact evaluation maps the measurement values to a value on an evaluation scale (e.g., school grades). *Quality Aspect Evaluations* provide rules for assessing quality aspects based on the evaluation results provided for (1) the impacts that influence the considered quality aspect and (2) its subordinated aspects. This means they aggregate several evaluation results into one result (e.g., by averaging or using a weighted sum).

## B. The Adaptation Process

We can distinguish three categories for QMs:

– *Public-level QMs* are intended for general use or use in a specific domain. Most of the models at this level are very generic; they are usually not operational and need to be customized. Using and tailoring these models could be useful for showing adherence to some standard.

– *Organization-level QMs* focus on satisfying the interests of a specific organization. They can focus on the whole organization, a business unit, or a project portfolio. They are intended to provide a common basis for project-specific model tailoring.

– *Project-level QMs* are applied to specify and assess quality for a specific project. Adaptation is limited to minor adjustments driven by project-specific requirements, without drastic changes to the organizational QM.

The general adaptation process we propose is applicable for adapting public models to obtain organizational models, refining an organizational model for business units or project portfolios, and deriving QMs addressing the needs of a specific project. The reuse potential is increased by means of step-wise refinement, which decreases the effort needed in further QM adaptations. The general adaptation process is illustrated in Fig. 2 and comprises four major steps:

*1. Specify goal of adapted QM*: The process begins by defining the goal of the QM that should result from the adaptation. To define this goal, the organization/project needs with respect to software quality and context information are used. In order to describe the goal in a structured way and not to forget important aspects, we use an adapted GQM goal template [3] with five goal parameters for the adapted model (GA), which is illustrated here with an example:

GA *Object* (i.e., considered artifact): Source code

*Purpose* (of the QM): Evaluation of product quality

*Viewpoint* (i.e., the perspective): User

*Focus* (i.e., the qualities of interest):

Reliability, Safety, Usability

*Context* (of planned model application):

Domain=Embedded; Language=Assembler

*2. Identify reference QM*: The goal is used to identify a model and adapt it to the needs of the project or organization. This model, on which the model adaptation is based, is called *reference model*. Finding the right reference model consists in finding the model whose goal parameters best fit the defined goal. In our example, we assume the following goal parameters characterizing the best fitting reference model (GR):

GR *Object*: Requirements specification, Source code

*Purpose*: Evaluation of product quality

*Viewpoint*: Developer, User

*Focus*: Maintainability, Reliability, Safety

*Context*: Dom.=Embedded; Paradigm=OO; Lang.=C, C++

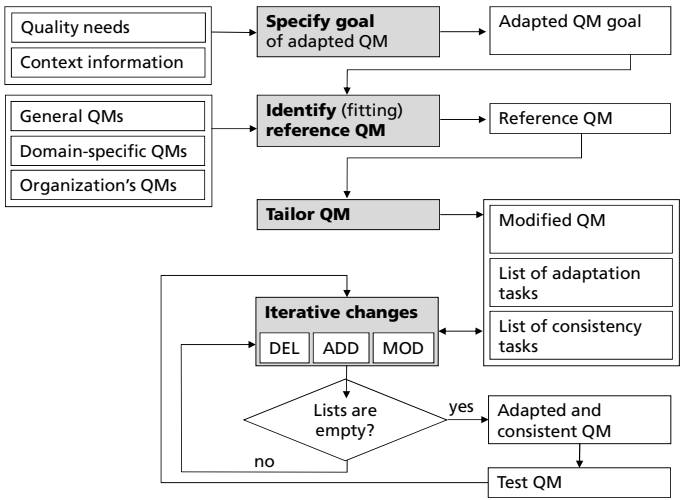

Figure 2.  Overview of the adaptation process with key activites (gray).

*3. Tailor QM*: Once a reference model is chosen, elements that are not needed in the final model are discarded. The unnecessary components are eliminated at the beginning in order to reduce the size and thus the complexity of the model. Sometimes, specific elements in the model can be reused in part but need some adjustments. During tailoring, such elements can be selected to stay in the model but are marked for detailed inspection and modification in the next step. Further, stubs for missing elements such as missing quality aspects can be added to provide reminders for needed model refinements in the next step. The adaptation rules used during the tailoring step make use of the goal parameters (GA/GR) and the structure provided by the Quamoco meta-model. They are summarized in Table I.

TABLE I.   TAILORING RULES (TR) FOR GOAL-BASED TAILORING

| |
|---|
| TR1: ∀ entity types ∉ GA.**object**: DEL(EntityType) |
| TR 2: ∀ elements of GA.object ∉ GR.object: ADD(EntityTypeStub) |
| TR3: IF(GA.**purpose**==specification): ∀ measures: DEL(Measure), ∀ impact evaluations: DEL(ImpactEval.), ∀ quality aspect eval.: DEL(QAspectEval.) |
| TR4: ∀ quality aspects ∉ GA.**viewpoint**: DEL(QualityAspect) |
| TR5: ∀ quality aspect eval. ∉ GA.viewpoint: DEL(QualityAspectEval.) |
| TR6: ∀ quality aspects not part of GA.**focus**: DEL(QualityAspect) |
| TR7: ∀ elements of GA.focus ∉ GR.focus: ADD(QualityAttributeStub) |
| TR8: ∀ factors not applicable based on their tags in GA.**context**: DEL(Factor) |
| TR9: ∀ measures not applicable based on tags in GA.context: DEL(Measure) |
| TR10: [opt.] add stubs for factor and measures introduced due to GA.context |

*Example:* For our goal definitions (GA) and (GR), the rules would propose and initiate the following actions: (TR1) Entities considering the requirements specification are removed since not needed. (TR4) All quality aspects and quality aspect evaluations considered only in the developer perspective are removed ("maintainability"). (TR7) All quality aspects not considered in the reference model ("usability") are added as dummies for further refinement. (TR9) Measures relevant only for "C" or "C++" (e.g., "depth of inheritance tree") and (PT8) factors relevant only for "object-oriented" programming (e.g., "documentation of classes") are removed.

(TR10) Stubs for measures relevant for assembler code should be added.

*4. Iterative Changes*: After sorting out irrelevant information, the model obtained might not be consistent or operational anymore. Therefore, the actions performed during the tailoring (e.g., removal of model components) triggers further consistency and adaptation tasks. These tasks help to bring the model back to a consistent, operational state. Some tasks can be automated (*consistency tasks*). Other tasks will require user interaction, as they are based on user decisions (*adaptation tasks*). The remaining adaptation work can be performed incrementally by processing open adaptation tasks in a user preferred order. Completing a task may initiate further consistency and adaptation tasks (see Table II), since completing an adaptation task usually requires deleting (DEL), adding (ADD), or modifying (MOD) elements in the model. The extent to which these operations are used depends on the suitability of the reference model. Accomplishing all adaptation tasks will lead to a consistent model customized to the user's needs. At this point, the QM should be piloted to test its suitability for the specified application purpose.

TABLE II.   TYPE OF ITERATIVE QM CHANGES AND CONSEQUENCES

| Element & Op. | | Consequential Adaptation and Consistency Tasks<br>Consistency tasks can be automatically performed without user interaction<br>Adaptation tasks require an explicit user decision and are collected in a To-do list |
|---|---|---|
| Entity Type | DEL | ∀ associated factors: DEL(Factor) [C]<br>∀ subordinated entity types: DEL(EntityType) [C] |
| | ADD | "Set name and description of entity type" [MOD(EntityType)] [A]<br>Associate with 1 superordinate entity type [MOD(EntityType)] [A]<br>"Check which factors influencing the quality of interest can be built for entities of this type and create them." [ADD(Factor)] [A] |
| Factor | DEL | ∀ associated impacts: DEL(Impacts) [C]<br>IF(associated property is not used by other factor): "If the property is no longer needed, delete it." [DEL(Property)] [A]<br>IF( associated entity type is a leaf in its hierarchy AND is not used by other factors): "…, delete it." [DEL(EntityType)] [A] |
| | ADD | Associate with 1 property [A] and with 1 entity type [A]<br>IF(GA.purpose = = evaluation): Associate with ≥1 measure [A]<br>"Provide a description for the factor" [MOD(Factor)] [A]<br>"Define ≥1 impacts for the factor" [ADD(Impact)] [A] |
| | MOD | MOD(isQuantified): ∀ associated impacts: IF(impact has an impact evaluation): "Check that all relevant measures of the factor are associated with the impact evaluation." [MOD(ImpactEvaluation.uses)] [A] |
| Impact | DEL | ∀ associated impact evaluations: DEL(ImpactEvaluation) [C]<br>IF(associated factor has no other impacts): "If the associated factor is no longer needed, delete it." [DEL(Factor)] [A]<br>IF(associated quality requirement is not connected to other impact): "…, delete it." [DEL(QualityReq,)] [A]<br>IF( associated quality aspect is a leaf in its hierarchy AND is not influenced by any impact): "…, delete it." [DEL(EntityType)] [A] |
| | ADD | Associate with 1 QualityAspect [A], 1 QualityReq. [A], and 1 Factor [A]<br>"Set justification and effect of added impact" [MOD(Impact)] [A]<br>IF(GA.purpose == evaluation): Associate with 1impact eval. [A] |
| | MOD | MOD(isImpacted): IF(impact has an impact evaluation): "Check that all relevant measures of all associated factors are associated with the impact evaluation" [MOD(ImpactEval.uses)] [A] |
| Property | DEL | ∀ associated factors: DEL(Factor) [C] |
| | ADD | "Set name and description of property." [MOD(Property)] [A] |
| | | "Check which factors that influence quality in focus can be built with this property and add them." [ADD(Factor)] [A] |
| Quality Aspect | DEL | ∀ associated impacts: DEL(Impact) [C] |
| | | ∀ associated Q aspect eval.: DEL(QualityAspectEvaluation) [C] |
| | | ∀ subordinated quality aspects: DEL(QualityAspect) [C] |
| | ADD | "Set name and description of Q aspect." [MOD(QualityAspect)] [A] |
| | | Associate with 1 superordinate Q aspect [MOD(QualityAspect)] [A] |
| | | "Refine aspect with sub-aspects, if necessary." [ADD(QAspect)] [A] |
| | | IF(GA.purpose == evaluation): Associate with 1 Q aspect eval. [A] |
| | | "Check which factors influences the added aspect, add impact relationships for them." [ADD(Impacts)]. [A] |
| | MOD | MOD(QA.refinedBy): IF (evaluateBy!=null): "Assure that all Q aspect evaluations of sub-aspects refining the aspect are considered in the Q aspect evaluation." [MOD(QAspectEval.)] [A] |
| | | MOD(QA.influencedBy): IF (evaluateBy!=null): "Assure that all impact evaluations of impact influencing the aspect are considered in the Q aspect evaluation." [MOD(QAspectEval.)] [A] |
| Req | DEL | ∀ associated impacts: DEL(Impact) [C] |
| | ADD | "Set name and description of added Q req." [MOD(QReq.)] [A] |
| Measure | DEL | "Delete the measure from the evaluation rule of the impact evaluations that used it." [MOD(ImpactEvaluation)]. [A] |
| | ADD | "Provide name and measurement rule." [MOD(Measure)] [A] |
| | | Associate with ≥1 factor [A] and ≥1 impact evaluation. [A] |
| | MOD | MOD(measurement_rule): ∀ associated impacts: IF(impact evaluation exists): "Check that the modified measure is correctly used in the evaluation rule" [MOD(ImpactEval.)] [A] |
| Impact Eval | DEL | IF(GA.purpose == evaluation): "Delete associated impact or add new impact evaluation" [DEL(Impact)|ADD(ImpactEvaluation)] [A] |
| | ADD | Associate with 1 impact [A] and with ≥1 measure [A] |
| | MOD | MOD(uses): "Assure that the evaluation rule of the impact evaluation considers all used measures" [MOD(ImpactEval.)] [A] |
| Q Aspect Eval | DEL | IF(GA.purpose == evaluation): "Delete associated aspect or add new aspect evaluation" [DEL(QAspect)|ADD(QAspectEval.)] [A] |
| | ADD | Associate with 1 quality aspect [A] |
| | | "Provide an aggregation rule for the Q aspect evaluation that considers all evaluations of influencing impacts and subordinated quality aspects." [MOD(QAspectEval.)] [A] |

*Example:* We illustrate the iterative adaption of a small model excerpt (Fig. 3). Based on our tailoring example, one open task is to refine the stubs added for measures addressing assembler code (PT10). In this case we consider only one stub M1, which was added to quantify the factor F1 "Documentation of source code". Open tasks for the added measure M1 are (a) "Provide name and measurement rule." and (b) "Associate with ≥1 impact evaluation". In addition, for the impact evaluations IE1 and IE2 of F1, there are open tasks, since an association between F1 and M1 was defined [MOD(F1.isQuantified)]: (c) "Check that all relevant measures of the factor F1 are associated with the impact evaluation IE1." and (d) "Check that all relevant measures of the factor F1 are associated with the impact evaluation IE2." In a first step, we complete task (a) by providing a name for M1, e.g.,"% of documented assembler lines" and a fitting measurement rule that returns a value between 0 and 100% =>MOD(M1.measurement_rule). Since M1 is not associated with an impact evaluation, the corresponding rule does not create any new task. In a next step, we complete task (c) by associating the new measure of F1 with the impact evaluation IE1 =>MOD(IE1.uses). This also completes task (b) "Associate M1 with ≥1 impact evaluation" but results in a new adaption task (e) "Assure that the evaluation rule of the impact evaluation IE1 considers all used measures". We complete this task by specifying an evaluation rule for IE1 that maps the measurement results of M1 onto the evaluation scale. Task (d) can be completed in a similar way as task (c) and results in task (f) "Assure that the evaluation rule of the impact evaluation IE2 considers all used measures" which can be completed in a similar way as task (e). After completion, we have a consistent and complete model excerpt.

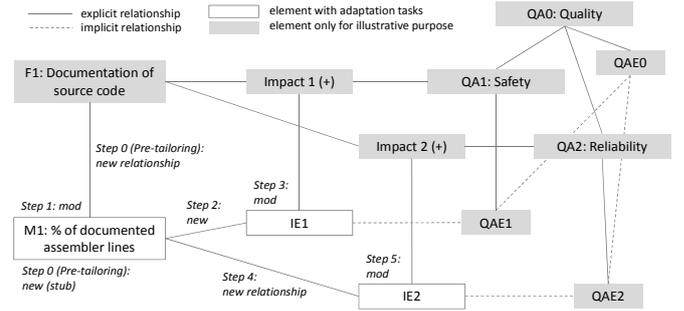

Figure 3. Sample QM excerpt to illustrate the iterative QM adaption

## IV. EMPIRICAL EVALUATION

In order to empirically evaluate the adaptation method, we compared QM adaptations performed *ad-hoc* using an existing *QM Editor* (E) against adaptations performed applying the *adaptation method* operationalized by an *Adaptation Assistant* (AA). The existing editor provides the capability to create and edit QMs but does not explicitly guide QM adaptations. The Assistant is implemented as a plug-in extending the Editor and supports the definition and comparison of QM goals, the tailoring, and updating of the list of open adaptation tasks.

### A. Study Goals

In the study, we wanted to investigate the quality of the proposed adaptation method; in particular, the following question was to be answered: '*Does the implemented adaptation method support the achievement of the three major requirements stated for quality model adaptation (R1-R3)?*' Thus, we defined three corresponding study goals:

*(G1) Formal Quality Model Consistency*: Evaluate whether the adaptation approach can improve the syntactical correctness of the adapted QMs. Consistent means the model conforms to the structure defined by the quality meta-model and a set of consistency rules. This means G1 addresses R1.

*(G2) Quality Model Appropriateness*: Evaluate whether the adaptation approach can improve the appropriateness of the adapted QMs. Appropriate means that the model is correct and complete with respect to its goal as specified during the specify goal activity (i.e., it is suitable for use with the object, purpose, viewpoint, quality focus and context), i.e., G2 addresses R2.

*(G3) Efficiency of Adaptation*: Evaluate whether the adaptation approach improves the efficiency of the adaptation. Efficient means the adaptation of the QMs can be performed in an effort-efficient manner. Consequently, G3 addresses R3.

### B. Study Context and Participants

The target population comprises people working as software quality managers in a company or in similar positions where part of their job is to adapt, set up, or maintain software QMs. We conducted the study in a workshop setting. The four participants were a mixture of practitioners and researchers experienced in working with QMs. In addition, they had experience with the Quamoco meta-model and the corresponding Editor. They had only rudimental knowledge regarding the proposed adaptation method and no experience with the Adaptation Assistant. To prepare the participants for the study, we presented the adaptation method together with brief examples. After that, we introduced its implementation provided by the Adaptation Assistant add-on.

### C. Concept Operationalization

We collected subjective judgments to investigate the three major study goals by asking the study participants closed questions related to the goals. Each question had to be answered on a 7-point *Likert scale*: {1: strongly disagree … 7: strongly agree} plus the answer option "I don't know".

– *Perceived_consistency:* "Do you consider the QM obtained to be syntactically correct?" This is a subjective assessment by the participants of formal QM consistency (G1).
– *Perceived_appropriateness:* "Do you consider the QM obtained to be appropriate with respect to its goal (i.e., the model is complete and correct with respect to its goal)?" This is a subjective assessment of appropriateness (G2).
– *Perceived_efficiency:* "Do you think that the adaptation can be performed efficiently?" This is a subjective assessment of the efficiency of the adaptation (G3).

Besides evaluating the goals based on the perception of the participants, we also wanted to evaluate them in a more objective way. Since it is difficult to objectively determine the degrees to which R1 and R2 are fulfilled directly, we addressed them indirectly by identifying the minimum set of model elements that need to be adapted (i.e., added, modified, or deleted) in order to obtain a consistently and appropriately adapted QM. This allows us to define measures regarding the completeness and correctness of the performed adaptation and use the measurement results as a more objective indicator for the model's consistency and appropriateness: A more completely and correctly adapted model is more consistent and appropriate.

*Completeness*: We say that a QM is completely adapted if all of its elements are adapted that needed to be adapted to obtain a model that is consistent with the structure described by the meta-model and appropriate for addressing its goal. We measure this concept using two base measures: the total number of elements that should be adapted in the QM based on the provided adaptation scenario and the number of elements in the QM that were adapted by the study participant:

$$\text{completeness} = \frac{\text{number of adapted elements that should be adapted}}{\text{number of elements that should be adapted}}$$

*Correctness*: We say that a QM is correctly adapted if all of its elements that need to be adapted are correctly adapted with respect to the goal of the adapted QM and defined consistency rules. This means that we measure the degree of correctness as the percentage of correctly adapted elements:

$$\text{correctness} = \frac{\text{number of correctly adapted elements}}{\text{number of elements that should be adapted}}$$

*Efficiency*: We measured efficiency in a more objective way by relating the number of correctly adapted elements and the time needed for the adaptation:

$$\text{efficiency} = \frac{\text{number of correctly adapted elements}}{\text{time required for adaptation}}$$

### D. Hypotheses

During the study, we tested the following six hypotheses:

**H$_{Sub1}$** (*Perceived consistency*): The participants consider the QMs obtained using the Adaptation Assistant (AA) to be more correct syntactically than the QMs obtained using the Editor (E):

$$\mu(\text{perceived\_consistency}(AA)) > \mu(\text{p\_con}(E))$$

**H$_{Sub2}$** (*Perceived appropriateness*): The participants consider the QMs obtained using the Adaptation Assistant to be more complete and correct with respect to their goals than the QMs obtained using the Editor:

$$\mu(\text{perceived\_appropriateness}(AA)) > \mu(\text{p\_app}(E))$$

**H$_{Sub3}$** (*Perceived efficiency*): The participants consider the adaptation to have been more efficiently performed using the Adaptation Assistant than using the Editor:

$$\mu(\text{perceived\_efficiency}(AA)) > \mu(\text{p\_eff}(E))$$

**H$_{Cmp}$** (*Completeness*): The adapted QMs obtained using the Adaptation Assistant are more completely adapted than the adapted QMs obtained using the Editor:

$$\mu(\text{completeness}(AA)) > \mu(\text{completeness}(E))$$

**H$_{Crr}$** (*Correctness*): The adapted QMs obtained using the Adaptation Assistant are more correctly adapted than the adapted QMs obtained using the Editor:

$$\mu(\text{correctness}(AA)) > \mu(\text{correctness}(E))$$

**H$_{Eff}$** (*Efficiency*): QM adaptation is more efficiently performed when using the AA than when using the Editor:

$$\mu(\text{efficiency}(AA)) > \mu(\text{efficiency}(E))$$

### E. Study Design and Implementation

In the study, each participant assumed the role of a quality manager and was asked to perform the following activities:

– *Finding most suitable reference model:* The participants had to select a reference model from a pool of QMs based on a provided adaptation scenario. Most suitable means that the model meets most of the characteristics requested.
– *Producing an adapted QM:* The participants had to execute adaptation tasks based on a provided adaptation scenario.

TABLE III.    STUDY DESIGN

|  | QM Editor | Adaptation Assistant |
|---|---|---|
| Group 1* | Adaptation Scenario A | Adaptation Scenario B |
| Group 2* | Adaptation Scenario B | Adaptation Scenario A |

*Both groups had the same number of randomly assigned participants.

Both activities were performed by the four participants twice: once with one scenario and the Editor and once with a second scenario and the Adaptation Assistant. We chose a cross-design with two different adaptation scenarios (Table III) in order to deal with the low number of participants while keeping the design-inherent learning effects low. After each adaptation scenario, the participants provided their feedback by filling out a questionnaire, which asked them to subjectively rate the formal consistency and appropriateness of the obtained QM as well as the efficiency of the adaptation. After the execution of both scenarios, the entire work-space of each participant was collected and saved for subsequent analysis. Based on this analysis, the completeness, correctness, and efficiency values were determined.

For the study, we provided each participant with the following input: (1) two QM application goals that should be used by the participants to find the most appropriate reference model, (2) two pools of QMs from which the most appropriate reference model should be selected by the participants on paper and in the adaptation tool, (3) two adaptation scenarios including practical adaptation task descriptions, (4) two example QMs that should be adapted by the participants.

*F. Study Results*

In this subsection, we present the descriptive statistics for the variables measured and the results of hypotheses testing.

*Descriptive Statistics*: Table IV shows the mean, median, and standard deviations (stdev) for the eight adaptations performed during the study, separated into applications of the Editor (our baseline) and the Adaptation Assistant.

TABLE IV.    STUDY RESULTS

|  | QM Editor | | | Adaptation Assistant | | |
|---|---|---|---|---|---|---|
|  | mean | median | stdev | mean | median | stdev |
| Completeness (in %) | 15.00 | 15.78 | 6.76 | 78.55 | 76.52 | 7.46 |
| Correctness (in %) | 8.93 | 9.98 | 3.87 | 70.34 | 69.17 | 8.82 |
| Efficiency (elements/min) | 0.37 | 0.41 | 0.18 | 2.90 | 2.84 | 0.54 |
| Perceived Consistency* | 3.50 | 3.50 | 2.38 | 6.00 | 6.00 | 0.82 |
| Perceived Appropriateness* | 2.00 | 2.00 | 0.82 | 5.75 | 6.00 | 0.50 |
| Perceived Efficiency* | 1.25 | 1.00 | 0.50 | 5.75 | 5.50 | 0.96 |

*measured using a 7-point Likert scale with 1: strongly disagree, 2: disagree, 3: somewhat disagree, 4: neither agree nor disagree, 5: somewhat agree, 6: agree, 7: strongly agree.

*Hypotheses*: As our sample was not large enough to assume a normal distribution, we applied non-parametric one-sided Wilcoxon signed-rank tests with alpha=0.05.

– $H_{Sub1}$ (perceived_consistency): *accepted* (p=0.032)
– $H_{Sub2}$ (perceived_appropriateness): *accepted* (p=0.033)
– $H_{Sub3}$ (perceived_efficiency): *accepted* (p=0.034)
– $H_{Cmp}$ (completeness): *accepted* (p=0.034)
– $H_{Crr}$ (correctness): *accepted* (p=0.034)
– $H_{Eff}$ (efficiency): *accepted* (p=0.034)

*G. Threats to Validity*

The two major threats to the validity of our results are the small sample size and the potential learning effects.

*Convenience sample and sample size*: The participants were chosen due to their experience in quality modeling in general and with the quality meta-model as well as with the Editor in particular. Therefore, there were only a limited number of potential participants, resulting in a convenience sample of limited size. However, the participants are more representative of the target population (i.e., professionals performing QM adaptations as part of their job) than, for example, graduated students of computer science or software engineering.

*Potential learning effects*: Although the participants were not requested to follow a particular process for adapting the first model using the Editor and were confronted with two different adaptation scenarios, they may have learned from the first adaptation, which may have positively influenced their performance during the second adaptation using the Assistant.

Further threats are that only a limited timeframe was available for the participants to conduct the adaptation tasks and that the attitude of the participants toward the well-known Editor or the newly introduced Adaptation Assistant may have influenced their subjective evaluation result.

*H. Interpretation*

Not only could all stated hypotheses be accepted, but the magnitude of the improvement using the tool-supported adaptation method also seems to be high when compared to performing the adaptation without explicit adaptation support using only the Editor. Completeness and correctness could be improved by an average of ~60%. The efficiency of the adaptation could be increased by ~factor 8. Moreover, the effect was perceived by the participants and could be measured by analyzing the adapted models. Therefore, although several threats to the study's validity exist, we conclude that the proposed adaptation method can increase the efficiency of adaptation tasks and the quality of their results in terms of consistent and appropriate models. Further, the study results indicate that typical QM adaptations are difficult to handle adequately without a tool-supported adaptation method. The main reason for these results appears to be that even at first glance, manageable adaptation tasks result in many subsequent sub-tasks that must be performed in order to assure the completeness and correctness of the adapted model. In part, these sub-tasks are hard to identify without support due to the complexity of a typical QM, and even harder to remember until they can be resolved due to their large number, especially if there is no process providing guidelines throughout the adaptation.

## V. SUMMARY AND FUTURE WORK

We illustrated that adapting models is important for getting QMs that fit the needs of a concrete application context without building each model from scratch. However, in many cases the adaptation of a QM is a complex and error-prone task.

Therefore, we presented a flexible but rigorous approach to adapting QMs under the assumption that they conform to a principal structure provided by an appropriate meta-model. The proposed method addresses the need for efficiently adapting QMs in a way that results in consistent and appropriate models. The consistency of the adapted QM is covered by the definition of elementary adaptation operations and corresponding consistency rules; further, the method integrates a structured definition of the QM goal and addresses the efficiency of adaptation through automation (goal-oriented) tailoring.

The conducted study indicates that the performance of a QM adaptation can be significantly improved when using a well-defined and tool-supported adaptation method such as the one presented in this paper. Not only were the consistency and appropriateness of the adapted QM significantly improved, but so was the efficiency of performing the adaptation tasks.

In a next step, the adaptation method including the rules for identifying the required adjustment tasks should be transferred to an updated QM structure and get evaluated in an industrial field study in order to ensure its applicability in practice.

ACKNOWLEDGMENT

Parts of this work have been funded by the BMBF project Quamoco (grant no. 01IS08023C). We gratefully acknowledge Jens Göddel for his contributions and thank Sonnhild Namingha for reviewing a first version of this article.